\documentclass[11pt,twoside]{article}

\usepackage{asp2006}
\usepackage{epsf}
\usepackage{lscape}

\begin{document}
\title{Compact starburst in the central regions of Seyfert galaxies}   
\author{Kotaro Kohno}   
\affil{2-21-1 Osawa, Mitaka 181-0015, Tokyo, Japan, 
Institute of Astronomy, University of Tokyo} 
\author{Koichiro Nakanishi, Masatoshi Imanishi}   
\affil{2-21-1 Osawa, Mitaka 181-8588, Tokyo, Japan, 
National Astronomical Observatory of Japan}   

\begin{abstract} 
We have conducted
a high-resolution ``3D'' imaging survey of the CO(1--0), HCN(1--0), and HCO$^+$(1--0) lines toward the 
central a few kpc regions of the Seyfert and starburst galaxies in the local universe
using the Nobeyama Millimeter Array.
We detected luminous HCN(1--0) emissions toward a considerable fraction of these
Seyfert galaxies (10 of 12 in our sub-sample), which indicated that some of these Seyfert galaxies,
such as NGC 3079, NGC 3227, NGC 4051, NGC 6764, and NGC 7479, are
indeed accompanied with compact nuclear starburst, given the tight correlation 
between the HCN(1--0) luminosity and the star formation rate among star-forming galaxies.
However, we suggest that the elevated HCN(1--0) emission from some of these Seyfert galaxies,
including NGC 1068, NGC 1097, NGC 5033, and NGC 5194, does not signify the presence
of massive starbursts there. 
This is because these Seyfert nuclei show abnormally high HCN(1--0)/HCO$^+$(1--0)
ratios (2--3), which were never observed in the starburst nuclei in our sample. 
This could be attributed to the overabundance of HCN molecules in the X-ray dominated regions (XDRs)
at the centers of these Seyfert galaxies.
\end{abstract}


\section{HCN as a tracer of star formation in galaxies}   

A dense molecular medium plays various roles in the
vicinity of active galactic nuclei (AGNs). The presence of spatially compact
dense and dusty interstellar matter (ISM), which obscures the
broad-line regions in the AGNs, is inevitable 
according to the proposed unified model of Seyfert
galaxies. This circumnuclear dense
ISM could be a fuel reservoir for active nuclei as well as a site for 
massive star formation. In fact, a strong enhancement of HCN(1--0)
emission with respect to CO has been detected in the prototypical
type-2 Seyfert NGC 1068 \citep{jackson1993, tacconi1994, helfer1995}.
Similar enhancements have also been reported
in NGC 5194 \citep{kohno1996},
NGC 1097 \citep{kohno2003}, and NGC 5033 \citep{kohno2005}.
In these Seyfert nuclei, the HCN(1--0) to CO(1--0) integrated intensity ratios in the 
brightness temperature scale, $R_{\rm HCN/CO}$, are enhanced up to approximately 0.4--0.6, 
and the kinematics of the HCN line indicates that this dense molecular medium could be 
the outer envelope of the obscuring material 
\citep{jackson1993,tacconi1994,kohno1996}.

On the other hand, 
it is well known that there exists a {\it tight} and {\it linear} correlation between HCN(1--0) 
and FIR luminosities among star-forming galaxies in the local universe
\citep{gao2004a}.
Therefore, one may immediately wonder if 
massive star formation occurs at the very centers of these Seyfert galaxies.
Is HCN emission still a tracer of massive star formation there? 

To answer this question, we attempted to find hints from our
high-resolution ``3D'' imaging survey of CO(1--0), HCN(1--0) and HCO$^+$(1--0) lines toward the 
central a few kpc regions of the Seyfert and starburst galaxies
using the Nobeyama Millimeter Array \citep{kohno2001,kohno2005}. 
This paper provides a brief summary of the current survey results 
and their implications on the presence of compact nuclear starbursts in nearby
Seyfert galaxies.

\section{Nobeyama Millimeter Array Imaging Survey of CO, HCN, and HCO$^+$ emissions 
toward Seyfert and starburst galaxies}

The majority of the Seyfert sample galaxies belong to the Palomar Northern Seyfert sample
\citep{ho2001}.
Some southern Seyfert galaxies are also included in this sample. 
The band width of correlator, 1 GHz \citep{okumura2000}, enables us 
to detect the HCN(1--0) and HCO$^+$(1--0) lines simultaneously. 
This allowed us to make accurate measurements of the ratios of their relative line intensities,
i.e., $R_{\rm HCN/HCO^+}$.
Our HCN and HCO$^+$ cubes have typical resolutions of $\sim 2''$ to $6''$ 
(or a few 100 pc) and 
sensitivities of a few mJy beam$^{\rm -1}$
for a $\sim$50 km s$^{\rm -1}$ velocity channel.
We detected luminous HCN(1--0) emission toward a considerable fraction of these
Seyfert galaxies (10 of 12 Seyfert galaxies in our sub-sample). 
Among them, we present the molecular line images
of the two type-1 Seyfert galaxies NGC 1097 and NGC 7469 in figure 1.

To investigate whether or not these HCN and HCO$^+$ emissions indeed trace massive star formation in these regions,
we computed the ratios, $R_{\rm CO/HCN}$ and $R_{\rm HCN/HCO^+}$.
Some of the Seyfert galaxies, including NGC 1068, NGC 1097,
NGC 5033, and NGC 5194, show enhanced 
or overluminous HCN emission with respect to the CO and HCO$^+$ emissions.
The $R_{\rm CO/HCN}$ and $R_{\rm HCN/HCO^+}$ ratios
in these Seyfert galaxies are
enhanced up to $\sim$0.2--0.3 and $\sim$2--3, respectively.
Crucially, such elevated $R_{\rm HCN/HCO^+}$ values were {\it never} observed 
in our nuclear starburst sample.
Given the similar properties of these two molecules (i.e., similar permanent
dipole moments and therefore similar critical densities for collisional excitation),
the enhancement of $R_{\rm HCN/HCO^+}$ close to or larger than 2 is unusual.
Note that we found no clear correlation between $R_{\rm HCN/CO}$ and the morphologies
of the host galaxies.

\begin{figure}[!ht]
\plottwo{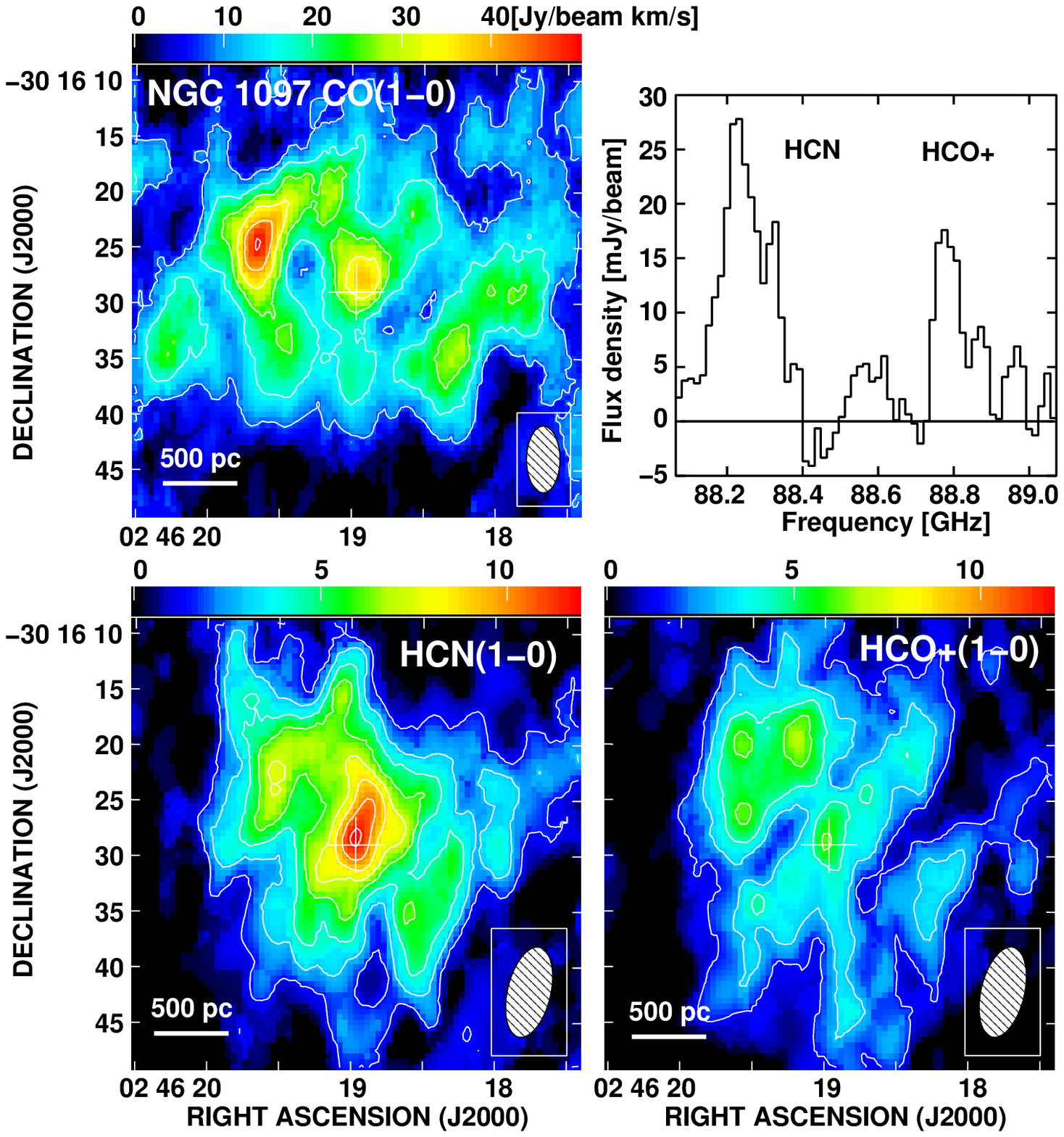}{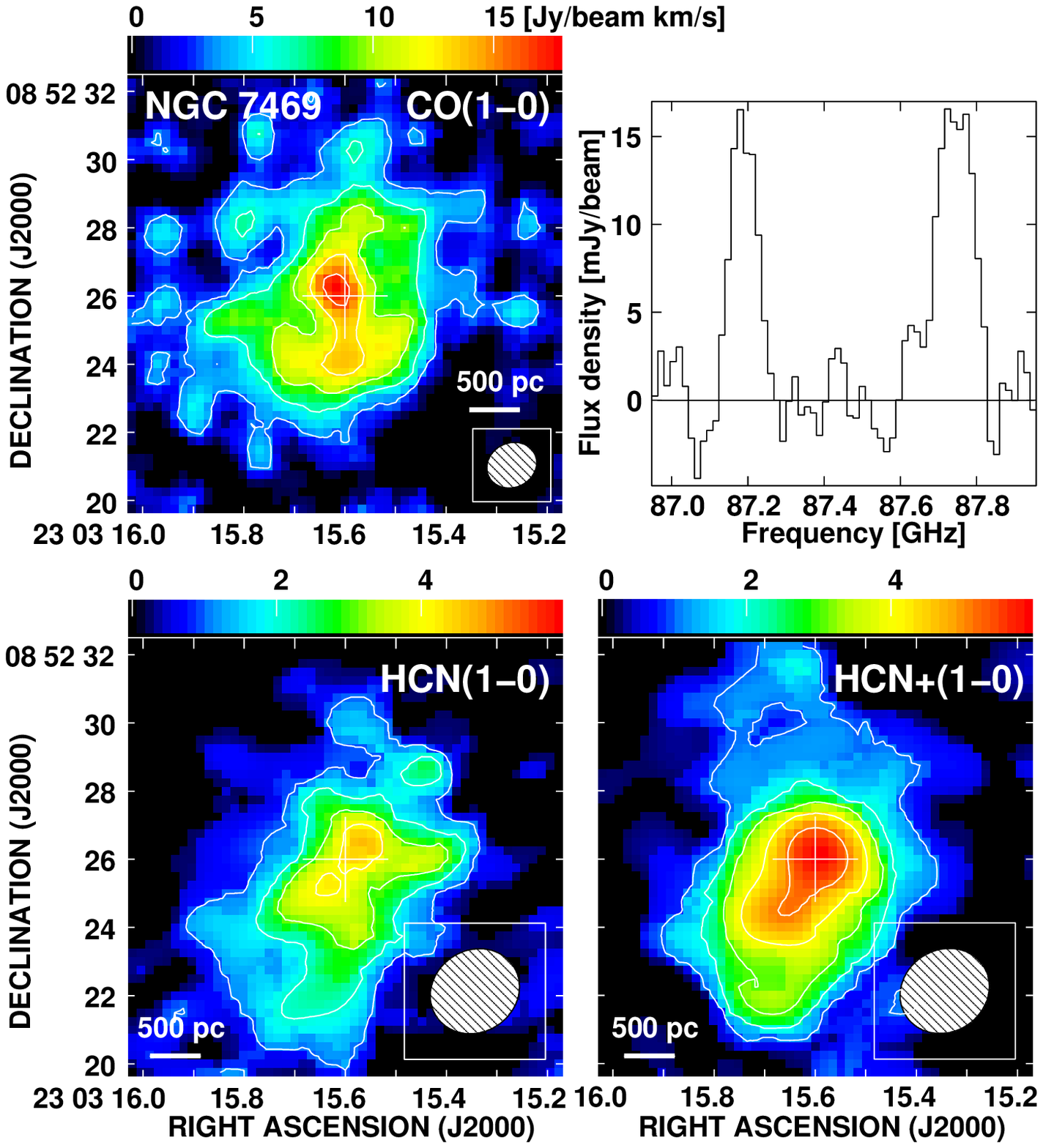}
\caption{
CO(1--0), HCN(1--0), and HCO$^+$(1--0) images of NGC 1097 (left) and NGC 7469 (right).
HCN(1--0) and HCO$^+$(1--0) spectra at the nuclei positions are also shown.
In NGC 1097, the enhancement of HCN(1--0) emission with respect to other emissions is evident;
HCN(1--0) emission is dominated by the nucleus, whereas CO(1--0) and HCO$^+$(1--0) emissions
are more luminous in the circumnuclear starburst ring ($r \sim 10''$).
}
\end{figure}

One possible explanation for these abnormally luminous HCN emissions
with respect to the CO and HCO$^+$ emissions is the chemistry due to X-ray dominated regions (XDRs)
\citep{maloney1996}, 
i.e., the overabundance of the HCN molecules in X-ray irradiated dense molecular tori
\citep{lepp1996,meijerink2005}.
One of the key issues is the high temperature of the molecular clouds in XDRs;
in contrast to the photo-dissociation regions (PDRs), where
UV photons are blocked at the surface of molecular clouds,
high-energy photons can penetrate deep inside molecular clouds.
Besides, heating due to photo-ionization in the XDRs is much more efficient than
the photo-electric heating in the PDRs. As a consequence,
the temperature of the molecular clouds in the XDRs becomes very high
as compared with that of the molecular clouds in the PDRs \citep{maloney1999,meijerink2005}.
In fact, at the center of M 51, a host of low-luminosity AGN
(see references in \cite{kohno1996}), a very high kinetic temperature
of the molecular gas has been suggested \citep{mat1998,mat2004}.
This nucleus is a representative one that shows the overluminous HCN(1--0) emission
in our sample ($R_{\rm HCN/HCO^+} = 2.5 \pm 0.43$).

Our interpretation is also supported by the comparison of our results 
with these obtained by infrared $L$-band spectroscopy:
a polycyclic aromatic hydrocarbon (PAH) emission at 3.3 $\mu$m
in the $L$-band can be considered as a good probe to study
nuclear starbursts in Seyfert galaxies \citep{imanishi2002, imanishi2003}.
In table 1, our diagnostic results on the presence of nuclear starburst based on the $R_{\rm HCN/HCO^+}$ values 
were compared with those from PAH observations.
These two diagnostics provided us with the same conclusions in 6 of 7 Seyfert galaxies.
We may need to further investigate the significance of the disagreement in the nucleus
of NGC 7469; one possibility is that our spatial resolution is still not sufficient to eliminate
the contamination from the circumnuclear starburst regions of NGC 7469.

\begin{table}[!ht]
\caption{Presence of nuclear starburst in nearby Seyfert galaxies: 
diagnostics from HCN/HCO$^+$ ratio and PAH emission}
\smallskip
\begin{center}
{\small
\begin{tabular}{lccl}
\tableline
\noalign{\smallskip}
Name &  \multicolumn{2}{c}{Nuclear starburst?} & Ref.\ for PAH \\
\noalign{\smallskip}
\cline{2-3}
\noalign{\smallskip}
  & HCN/HCO$^+$ & PAH &   \\
\noalign{\smallskip}
\tableline
\noalign{\smallskip}
NGC 1068 & No & No & \citet{imanishi2002} \\
NGC 3227 & Yes & Yes & \citet{imanishi2002}, \citet{rv2003} \\
NGC 4051 & Yes & Yes? & \citet{rv2003} \\
NGC 4388 & No & No & \citet{imanishi2003}  \\
NGC 4501 & No & No & \citet{imanishi2003} \\
NGC 5033 & No & No & \citet{imanishi2002} \\
NGC 7469 & Yes? & No? & \citet{iw2004} \\
\noalign{\smallskip}
\tableline
\end{tabular}
}
\end{center}
\end{table}

In summary, 
the known tight correlation between 
HCN(1--0) luminosities and SFRs in star-forming galaxies 
should be treated with caution in the vicinity of active nuclei,
provided the high $R_{\rm CO/HCN}$ and $R_{\rm HCN/HCO^+}$ values are indeed 
the signatures of HCN overabundance due to the XDR chemistry.
We suggest that {\it the overluminous HCN(1--0)
emission observed in some Seyfert galaxies such as NGC 1068, NGC 1097,
NGC 5033, and NGC 5194 does not signify an elevated massive star formation rate there}.
Nevetheless, it is still likely that compact nuclear starbursts occur in other Seyfert galaxies in our sample, 
such as NGC 3079, NGC 3227, NGC 4051, NGC 6746, and NGC 7479.
This is because the observed $R_{\rm CO/HCN}$ and $R_{\rm HCN/HCO^+}$ values 
in these galaxies are very similar to those of nuclear starburst galaxies.
The comparison of our results with those obtained from $L$-band PAH spectroscopy also seems to support our conclusions.

\acknowledgements 

The authors would like to thank the collaborators, including T.~Shibatsuka, M.~Okiura,
T.~Tosaki, T.~Okuda, S.~Onodera, M.~Doi, K.~Muraoka, A.~Endo, 
S.~Ishizuki, K.~Sorai, S.K.~Okumura, Y.~Sofue,
R.~Kawabe, and B.~Vila-Vilar\'o,
for their invaluable efforts.
We are grateful to the staff at NRO for operating the NMA.
A part of this study was financially supported by MEXT Grant-in-Aid for 
Scientific Research on Priority Areas No.~15071202.


\end{document}